\newcommand\setrow[1]{\gdef\rowmac{#1}#1\ignorespaces}
\newcommand\clearrow{\global\let\rowmac\relax}
\clearrow

\documentclass[journal]{IEEEtran}

\usepackage{makecell}

\usepackage{listings}
\usepackage{xcolor}

\definecolor{codegreen}{rgb}{0,0.6,0}
\definecolor{codegray}{rgb}{0.5,0.5,0.5}
\definecolor{codepurple}{rgb}{0.58,0,0.82}
\definecolor{backcolour}{rgb}{0.95,0.95,0.92}

\usepackage{cite}
\usepackage{caption}% 

\ifCLASSINFOpdf
  \usepackage[pdftex]{graphicx}

  \graphicspath{{../pdf/}{../jpeg/}}

  \DeclareGraphicsExtensions{.pdf,.jpeg,.png}
\else

  \usepackage[dvips]{graphicx}

  \graphicspath{{../eps/}}
  \DeclareGraphicsExtensions{.eps}
\fi

\usepackage{amsmath,amssymb,amsfonts}
\usepackage{algorithmic}
\usepackage{array}
\usepackage{comment}
\usepackage{fixltx2e}
\usepackage{stfloats}
\usepackage{flushend}

\usepackage{enumitem}

\usepackage{algorithm2e}
\usepackage{xcolor}

\newcommand{\xvar}[1]{\textsf{#1}}

\newcommand{\xvbox}[2]{\makebox[#1][l]{#2}}

\SetAlFnt{\small} %\footnotesize

\SetAlCapSty{xAlCapSty}

\SetCommentSty{xCommentSty}

\SetNlSty{mynlfont}{}{} %\scriptsize

\LinesNumbered

\SetSideCommentRight

\DontPrintSemicolon

\RestyleAlgo{algoruled}

\usepackage[underline=true,rounded corners=false]{pgf-umlsd}
\definecolor{cor-very-weak}{HTML}{BBBBBB}
\definecolor{cor-weak}{HTML}{EEBD84}
\definecolor{cor-moderate}{HTML}{F47461}
\definecolor{cor-strong}{HTML}{F47461}
\definecolor{cor-very-strong}{HTML}{8B0000}

\begin{document}
\title{Blockchain based AI-enabled Industry 4.0 CPS Protection against Advanced Persistent Threat} 

\begin{comment}
\markboth{IEEE Journal of Internet of Things, February~2022}%
{Shell \MakeLowercase{\textit{et al.}}: Protecting Industry 4.0 CPS from Advanced Persistent Threat (APT) using Consortium Blockchain and Edge-complaint Reusable Machine Learning Techniques }

\author{
    \IEEEauthorblockN{Ziaur Rahman, Xun Yi,  Ibrahim Khalil\\}
    \IEEEauthorblockA{School of Computer Science and Software Engineering
    \\RMIT University, Melbourne, Australia\\\{rahman.ziaur, xun.yi, ibrahim.khalil\}@rmit.edu.au}
}

\IEEEpubid{0000--0000/00\$00.00~\copyright~2021 IEEE}

\maketitle

\end{comment}

\author{Ziaur Rahman, Xun Yi, and Ibrahim Khalil % <-this % stops a space

\thanks{Manuscript received September 15, 2021; revised December 16, 2021; accepted January 19, 2022. Date of publication February , 2022; date of current version January, 2022. The work was supported by RMIT Research Stipend Scholarship (RRSS) Program. The work of Xun Yi was supported in part by the Project 
``Privacy-Preserving Online User Matching'' under the grant ARC DP180103251. Corresponding author: Ziaur Rahman.}

\thanks{Ziaur Rahman, Xun Yi, and Ibrahim Khalil are with the School of Science, RMIT University, Melbourne, Australia. e-mail: (rahman.ziaur, xun.yi, ibrahim.khalil)@rmit.edu.au}
\thanks{Digital Object Identifier: 10.1109/JIOT.2022.XXXXX}

% \IEEEpubidadjcol

%\thanks{Mehdi Sookhak and Shaoen Wu are with School of Information Technology, Illinois State University, Normal, IL, USA. e-mail: m.sookhak@ieee.org, swu1235@ilstu.edu.}

%\thanks{Mehdi Sookhak is also with Dept. of Computer Science, Texas A\&M University-Corpus Christi, 6300 Ocean Drive, Corpus Christi, Texas, USA, 78412.}

}


\markboth{IEEE INTERNET OF THINGS JOURNAL,~Vol., No., February~2022}%
{Ziaur \MakeLowercase{\textit{et al.}}: Bare Demo of IEEEtran.cls for IEEE Communications Society Journals}



\maketitle

\IEEEpubidadjcol
%%%%%%%%%%%%%%%%%%%%%%%%%%%%%%%%%%%%%%%%%%%%%%%%%%%%%%%%%%%%%%%%%%%%%%%%%%%%%%%%

\begin{abstract}

Industry 4.0 is all about doing things in a concurrent, secure, and fine-grained manner. IoT edge-sensors and their associated data play a predominant role in today's industry ecosystem. Breaching data or forging source devices after injecting advanced persistent threats (APT) damages the industry owners' money and loss of operators' lives. The existing challenges include APT injection attacks targeting vulnerable edge devices, insecure data transportation, trust inconsistencies among stakeholders, incompliant data storing mechanisms, etc. Edge-servers often suffer because of their lightweight computation capacity to stamp out unauthorized data or instructions, which in essence, makes them exposed to attackers. When attackers target edge servers while transporting data using traditional PKI-rendered trusts, consortium blockchain (CBC) offers proven techniques to transfer and maintain those sensitive data securely. With the recent improvement of edge machine learning, edge devices can filter malicious data at their end which largely motivates us to institute a Blockchain and AI aligned APT detection system.  The unique contributions of the paper include efficient APT detection at the edge and transparent recording of the detection history in an immutable blockchain ledger. In line with that, the certificateless data transfer mechanism boost trust among collaborators and ensure an economical and sustainable mechanism after eliminating existing certificate authority. Finally, the edge-compliant storage technique facilitates efficient predictive maintenance. The respective experimental outcomes reveal that the proposed technique outperforms the other competing systems and models. 

%need improvement -zia

% While many initiatives have been taken globally to get optimal performance by these algorithms to prevent intrusion detection, it is still out of control.

\end{abstract}  

% Note that keywords are not normally used for peer-review papers.


% \fi
\IEEEpeerreviewmaketitle

\section{Introduction}

\IEEEPARstart{S}{ince} the last decade, the world has experienced the latest iteration of the industrial ecosystem called Industry 4.0. This fourth revolution demands the adoption of connected devices and techniques to meet the increasingly growing system protection requirements. The ultimate goals of building such automated connections range from enhancing productivity, reducing costs to boosting revenue. After merging advanced technology such as the Internet of Things (IoT), Artificial Intelligence (AI), etc., the latest industrial infrastructure has laid the foundation for the desired smart factory system, where the convergence happens between machines and humans depending on data. Data generated by edge sensors play a vital role in monitoring the manufacturing process, predicting maintenance, and detecting equipment anomalies. As a critical component, if data fails to comply with the security standard, all the actions associated with data will undoubtedly affect or paralyze the entire industrial ecosystem. The recent security issues published by the Guardian and ABC support the US and Australian claims and concern of stealing their industry copy-right data through cyber espionage by other countries. Even during the worldwide COVID 19 pandemic, Webber has recorded about 50 cyberattacks only in Australia since January 2020, which was 120 in the last two years.%\footnote{https://www.webberinsurance.com.au/}}%. 
The report shows that most attacks targeted large industries such as Bunnings, Alinta Energy, and Toyota, including sensitive health information. The country indulges a 15 billion dollar package in tackling potential threats and unprecedented loss. US Defense (DoD) also funded 8.5 billion in Cybersecurity, with an almost 5 percent increase over the previous year.
On top of the incidents above, security concern has become an inevitable issue that deserves proper addressing inside all processes of the Industry 4.0 system. However, the existing industrial security solutions are mostly designed, relying on the security mechanism in the server-side,  trust provided by the trusted third party (TTP) such as cloud and certificate service provider. 

\begin{figure}[htb!]
    \centering
    \includegraphics[width=\linewidth]{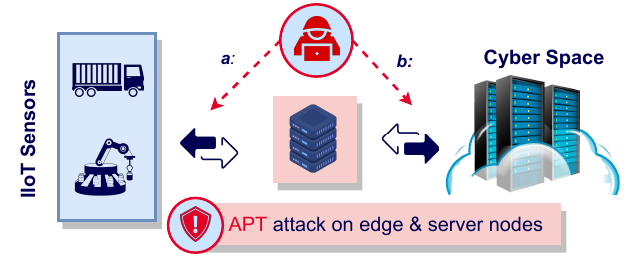}
    \caption{Cyberattacks to inject Advanced Persistent Threat (APT) to the Industry 4.0 CPS via a)  IoT Edge b) Cyberspace}
    \label{fig:fig_1}
\end{figure}

\subsection{APT Attack Model and Challenges}
Among several other intrusions and malwares injected to IoT edge nodes, the latest ransomware, namely Advanced Persistent Threat (APT) has caught broader attention because of its detective nature. As a stealthy threat actor, an APT strives to control a system network after remaining undetected over a long period. Though attackers target the server-side, currently, several incidents were recorded where the edge-side vulnerability was responsible. Therefore, today's Industry 4.0 network has to tackle that it has no APT inside the edge servers. As shown by Figure \ref{fig:fig_1}, APT may enter into the Industry 4.0 Cyber-physical System (CPS) both via edge and server nodes. There are several techniques that focus solely server-side protection \cite{mostafaDBF} \cite{BCTLWang}. Similarly some works focus edge protection using collaborative machine learning technique. Undeniably, the system can not be sustainable if there any  security loophole at the edge-end \cite{BCTLWang}. Existing approaches seem to be utilizing complex machine learning algorithms that requires significant computational capabilities which are often NOT edge complaint. Several works have used server-driven data to evaluate their proposed techniques which may not work at certain circumstances \cite{FedTLAccess}.

Advanced Persistent Threat (APT) is well funded, organized group that is systematically developed to compromise large-scale information of government and commercial entities. Malware is any malicious software or program designed to damage or disable computer systems or networks. APT is a broad term used to describe a prolonged, more strategic, and targeted attack. However, most malware attacks are target-specific, quick-damaging attacks. Besides, APT can stay undetected for an extended period; on the contrary, anti-malware tools can detect and eradicate malware from the system.

\subsection{Contributions}

With a motivation to protect both edge IoT and server-side data transfer the key contributions of the paper are as follows. 

\begin{itemize}
\item A blockchain based AI-enabled APT detection system is proposed that protect Industrial IoT data from being forged. 
\item Reusable machine learning method has been incorporated at the IoT edge that secure data before sending it to the cyberspace. 
\item Consrtium blockchain (CBC) brings trust among the participating stakeholders that prevent system from centralized dependency and facilitates sustainable system.
\item Certificateless device registration and data transfer teachnique has been proposed that save costs after eliminating certificate authority and brings collaborative operation
\item Immutable recording of both APT detection and data transaction in the blockchain ledger and storing in the edge-complaint distributed hash table (DHT) ensures higher performance and efficiency. Because of the DHT integration, the respective experimental outcomes reveal that the proposed technique outperforms the other system with competing machine learning models.
\end{itemize}

\subsection{Organisation}

The remainder of this article is organized as follows. Section II include the background and related state of the art literature. Section III explains the proposed model where the necessary evaluation is detailed through subsequent section IV. The final section conclude the future scopes and justify how authors achieves the claims made throughout the work.

\section{Background and Related Work}
Blockchain is a growing, publicly distributed, and permanent ledger to which transaction events are posted and verified by the peers on the network. The entire process happens without the intervein of any third-party, that makes it so appealing, indeed. Bitcoin is the most common example of Blockchain where data as transactions are maintained after being confirmed through an incentivized system in which members must compete to complete some proof-of-work like cryptographic challenge. One block is linked with its nearest block by using the hash of that block; therefore, any modification in the block breaks all the previous chain and consensus. The latest block establishes the integrity of the last block. 

\begin{figure}
    \centering
    \includegraphics[width=\linewidth]{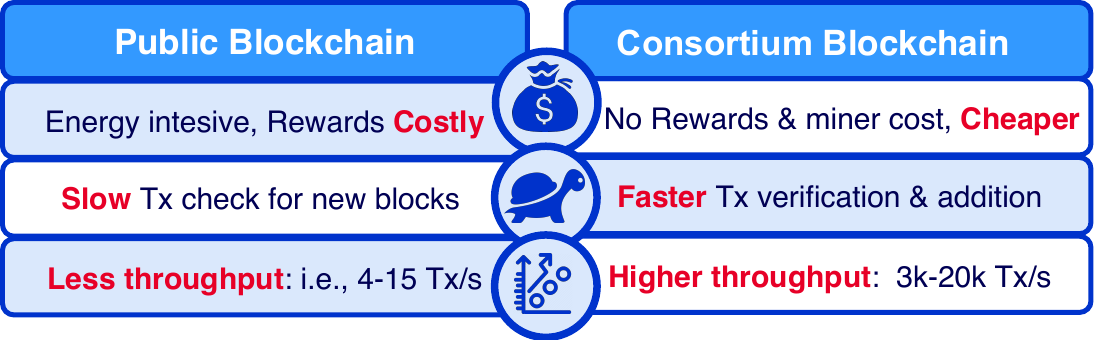}
    \caption{Consortium Blockchain suitability for Industrial IoT}
    \label{fig:fig_2}
\end{figure}

% \vspace*{-\baselineskip}

\subsection{Blockchain suitability for Industry 4.0 CPS}
Public blockchain best suits where an utterly untrusted network requires to be safe; however, it is slow and expensive. For example, for setting up a powerful mining node, in reality, is costly, on top of that, it requires enormous energy consumption to process the mining works. Besides, public blockchain can verify only a few transactions per second, which makes it incompatible for the use cases such as industry where plenty of data-transactions need to be done in real-time. On the other hand, consortium or permissioned type of blockchain such as Hyperledger (HLF), Quorum, Corda have a selective setup where only invited members instead of arbitrary participants are allowed to join the network who agreeably trust each other. Here token for incentives/rewards is not mandatory; thus, expenses for mining setup can be avoided to make it adaptable for the real-time and critical system like Industry 4.0 application. As participating nodes are acquainted beforehand, it brings natural protection against 'Sybil Attacks'. Figure \ref{fig:fig_2} shows that permissioned blockchain (BC) is cheaper, faster and also has higher tranaction processing rate. For example, Proof of Work (PoW)-driven BC, i.e. Ethreum has the rate of 4 to 5 transaction per seconf (Tx/s) where permissioned BC, i.e. HLF Fabric can process about 3,000 to 20,000 Tx/s which in essence make it a inevitable choice for the proposed industry 4.0 edge communication \cite{ZiaXComSoc}.

\subsection{Deep Transfer Learning}
\begin{figure}
    \centering
    \includegraphics[width=\linewidth]{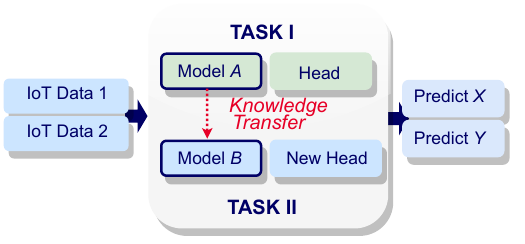}
    \caption{Knowledge transfer and prediction technique on different datasets in deep transfer learning (DTL) approach}
    \label{fig:fig_3}
\end{figure}

Deep Transfer learning (DTL) converges storing knowledge gained while solving one problem and applying it to another, i.e., the knowledge to detect malware top up the knowledge of the model that detects intrusion.  Figure \ref{fig:fig_3} depicts how a model transfers/reuses its knowledge to predict a decision in cooperation with another model performing different task. Deep transfer learning (DTL) is given in terms of domains and tasks. Suppose, a domain $\mathcal{D}$ consists of: a feature space $\mathcal{X}$ and a marginal probability distribution $P(X)$, with $X = \{x_1,...,x_n\}\in \mathcal{X}$. Let a domain, $\mathcal{D} = \{X, P(X)\}$, is an example of task with two elements. A label space $\mathcal{Y}$ and and objective predictive function $f:\mathcal{X} \rightarrow \mathcal{Y}$. $f$ predicts the respective label $f(x)$ of an instance $x$. This task, denoted by $\mathcal{T}=\{\mathcal {Y},f(x)\}$, can be obtained from the training dataset consisting of pairs $\{x_{i},y_{i}\}$, where $x_{i}\in X$ and $ y_{i}\in \mathcal {Y}$ \cite{eqnTL} \cite{ZiaShawon}. 

Assuming an input domain $\mathcal{D}_S$ and a training task $\mathcal{T}_S$, an output domain $\mathcal{D}_T$ as if $\mathcal{D}_S \neq \mathcal{D}_T$ OR $\mathcal{T}_S \neq \mathcal{T}_T$. DTL improves learning of the target predictive function $f_{T}(\cdot)$ in $\mathcal{D}_T$ using the knowledge in $\mathcal{D}_S$ and $\mathcal {T}_S$. Besides, how to store edge sensor data deserves illustrations.

\subsection{DHT for Edge Data Storage}
Storing data in the DHT and its associated pointer-address into blockchain ledger (BCL) best suits for the IoT edge, i.e., smart energy, implantable medical system, car, or any other industry 4.0 CPS, etc because of its salient features. In our proposed setup, when any external user asks for data access, thekey generating and distribution $(KGD)$ authenticates in cooperation with the required number nodes running the consortium blockchain network. It confirms the unique benefits such as traceability, accountability, removing trusted party, decentralized mechanisms, etc. over existing the cloud-driven centralized storage model. The efficient DHT adaption additionally makes the proposed technique robust, self-organizing, highly scalable, and fault-tolerant against different attacks, i.e, false query injection attacks, APT, Zeroday, etc. The proposed Industry 4.0 CPS data protection suits most of the DHT protocol, however, the demonstration integrates InterPlanetary File System (IPFS)%\footnote{https://ipfs.io/ and https://github.com/ipfs/dht-node} and Kademlia \footnote{https://docs.rs/kademlia-dht/1.2.0/kademlia\_dht}%
considering the fixed-sized routing, malware and APT attacks. Before storing data, both transaction and devices need to be authenticated.

\subsection{Certificateless Authentication}
Suppose there are three {$n=3$} parties namely Bob, Elen and Peter in an industry 4.0 setup who agreed to cosign their partial secret  $ps$ before registering a new device into the system. They are connected over a CBC and work as KGD authority. $KGD$ dissipates their public parameter with the connected Industry 4.0 IoT devices.

 The partial key is a concept in ordinary certificate-less authenticated encryption (CLAE) and identity-based encryption (IBE) that fixes the key-escrow problem. In the Hypeledger consortium blockchain setup, a built-in Membership Service Provider (MSP) works as a certificate authority (CA). The proposed technique replaces it with a PKI-like key generation centre (KGC), by adopting a Blockchain (BC) consortium collaborating with the Industry 4.0 CPS peers.

\paragraph{Key generation} Purposing to sign a $ps$ for a number of sensor devices, Bob, Elen and Peter agrees to pick  prime numbers $p$, $q$ and group generator $g$ (e.g. primitive root) as if $q \mid p$. Private and public key pair are the ring elements of $\mathbb{Z}_{p}$. Let their private keys are $x_1$, $x_2$ and $x_3$, therefore calculated public keys will be $y_i={g}^x_i$ where $( i=0,1,...,n)$ and aggregated public key $Y = \prod_{i=1}^{n} y_{i} \ mod \ p$. The key pairs will be $\{p,q,g,Y\},\{x_1,x_2,x_3\}$.

\begin{figure}[htb!]
    \centering
    \includegraphics[width=\linewidth]{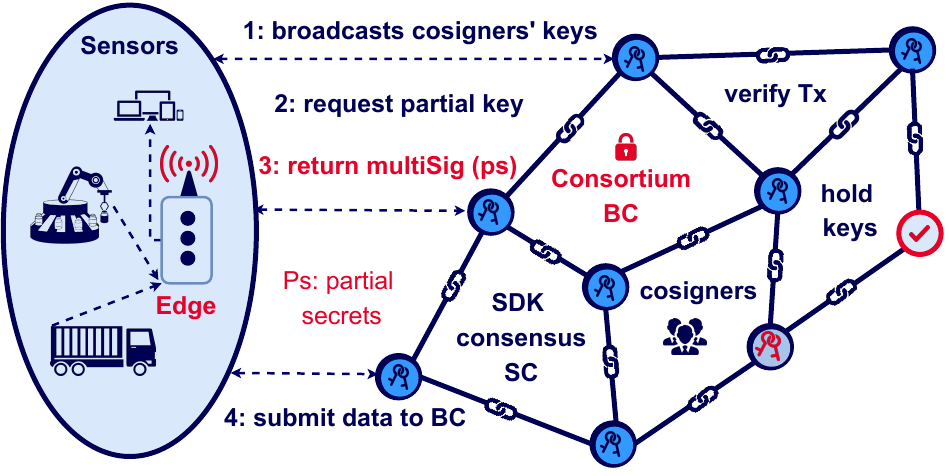}
    \caption{Certificateless communication between blockchain consortium (KGD) and Industry 4.0 IoT Edge devices}
    \label{fig:fig_CL}
\end{figure}

\paragraph{Signing data and partial secret} All cosigners choose random number $r_i$ such that {$ 0 < r < \mathbb{Z}_{q}$} and compute $R_1=g^r_i$ before finding the $R = \prod_{i}^{n} R_{i}\: mod \: p$. Suppose $T$ is the time-stamp including the dynamic edge identity $(E_{id})$ formed using  all individual sensor ids $(ID_j)$ and other required parameters, then the $KGD$ on Blockchain will find the signature parameter $c=H(T\parallel Y\parallel R\parallel{PS}$) here $H:M\rightarrow \mathbb{G}_{0}$ treated as random oracle in the security analysis and ${PS}$ are the KGD generated Partial Secrets (PS) for $m$ number of industrial IoT devices at a particular time $t$ that need to be multi-signed \cite{ZiaXComSoc}. Then the partial signature will be $s_i=(r_i+cx_i)\ mod \ q$ and the desired multi-signature will be $(R,S)$ where $S = \sum_{i=i}^{n} s_{i}\: mod \: p$ as demonstarted by Figure \ref{fig:fig_CL}.

\paragraph{Verifying device and data} The device receives the multi-signature $(R,S)$ along with the encrypted $ps$. As public key parameters such as $\{g,Y\}$ besides $T$ are already known to the edge sensors, it produces $c=H(T\parallel Y\parallel R\parallel{PS})$ using the same hash algorithm $H$. The device will accepts $ps$ before generating its own public-private key pairs $P_j,S_j$ if and only if it satisfies $g^S \ mod \ p = R \times Y^c \ mod \ p$.

\subsection{Related Work}
Blockchain immutable nature besides its pseudo-anonymity, traceability over the transparent distributed network have made Blockchain an unbeatable tool for Industry 4.0 CPS. Blockchain application found in the domain of copyright protection of digital data/asset, ID verification/provenance (notarization),  real state land ownership transfer, smart-taxation immigration, electronic voting, privacy-principle compliance (e.g., GDPR\cite{GDPR2019}).  Authors seemed to be practicing the immense benefits of distributed hash table (DHT) for storing access control and compliance data\cite{nameCoin2018}. However, besides the high-energy conducive miners' incentive disputes, the Blockchain network encounters the scalability issues that some existing-works concentrated on and aimed at solving through plausible remedies\cite{BCconsistency}\cite{lsbBC2019}\cite{rapidChain}. Considering the appealing features of Blockchain, researchers has incorporated AI (i.e. Deep learning, DTL, federated learning, etc.) with it. In an another work authors propose a cloud-based distributed deep learning framework for phishing and Botnet attack detection and mitigation \cite{DLBotnet}. A group of authors proposed a permissioned edge blockchain to secure the peer-to-peer (P2P) energy and knowledge sharing in framework to maximize edge intelligence efficiency \cite{enregyPermissioned}. Where \cite{mostafaDBF} proposes a deep blockchain framework (DBF) designed to offer security-based distributed intrusion detection and privacy-based blockchain with smart contracts in IoT networks.

%Similarly some recent works introduce edge computing into the Industrail IoT, so that the device can complete the federated learning operation \cite{FedTLAccess}.}%

One of the latest and motivating works proposed a consortium Blockchain based framework to protect Industry 4.0 CPS \cite{ZiaXComSoc}. There are a number of works studied addressed a novel blockchain-enabled model
sharing approach to improve the performance of
object detection with cross-domain adaptation for automatic driving systems \cite{drivingBC}. Authors addressed a special technique namely Authentication mechanism based on Transfer Learning empowered Blockchain, coined ATLB where blockchains are applied to achieve the privacy preservation for industrial applications \cite{BCTLWang}. Apart from this a group of researchers proposes a new transfer learning-based secure data fusion strategy (TSDF) for Industry 4.0 like system \cite{FusionTL}. Beside focusing reinforced machine learning 
scheme \cite{ReinforceBC} another group of authors proposed to enable Mobile Multi-user to make optimal offloading decisions based on blockchain transaction states, wireless channel
qualities. As studied, several mechanisms appear to have limitations to peer with the Industry 4.0 edge IoT protection, however, they have conceptually motivated us to design our proposed technique.

\section{Proposed APT Protection Mechanism}
\begin{figure}
    \centering
    \includegraphics[width=\linewidth]{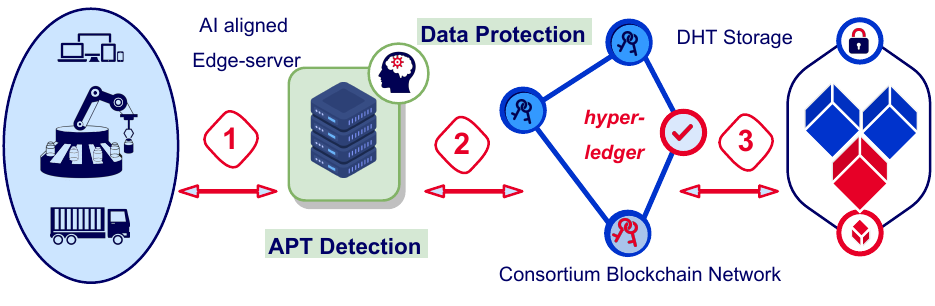}
    \caption{High Level view of the proposed blockchain based AI enabled Industry 4.0 CPS Data protection technique. It works within three steps. \textit{1)} Detecting APT at edge-server upon machine learning model \textit{2)} Consortium blockchain (CBC) validates none but the authentic sensors sending data and \textit{3)} Updating blockchain ledger (BCL) with the data transaction and APT detection status and finally storing data in the distributed hash table (DHT) storage. Connection establishment mechanism is illustrated in Figure 4 earlier.}
    \label{fig:fig_4}
\end{figure}

The protection scheme proposed here works in three steps. In the first step a deep transfer learning model gets deployed inside the edge server. The model is trained based on two combined and preprocessed datasets \cite{BotIotDS} \cite{TonIoTDS}  and the trained model is settled down in the edge server. Once edge server is called, it checks if there any advanced persistent threats found within that data. Secondly detection history along with the sensor data are sent to the linked DHT. Before storing the data it needs to check if there any APT injected during the data transfer over the network. In this step a blockchain consortium administers the process and ensure that only the registered and authenticated devices are sending data. In the final step blockchain smartcontract records the data transaction into the shared ledger and store data into the DHT storage. Figure \ref{fig:fig_4} portrays the steps one by one. The captions briefly shows those respectively.  

\subsection{Building DTL Model}
In order to identify the problems in the IoT environment better, Table \ref{Table: Symbols and description} gives the topical symbols and descriptions, in which we set the initial model that has enough labeled data to build an efficient intrusion detection model. When the new complex type of cyber-attacks arrives, the detection model is suitable for the new type of cyber-attacks \cite{ZiaRubel}. 

\textbf{Source Domain:} The domain where the initial model is located. The source domain data ($D_s : (X_s,Y_s)$) is the combination of $ {(X_{s1},Y_{s1}), (X_{s2},Y_{s2}), (X_{s3},Y_{s3}), ...... (X_{sn},Y_{sm})}$, in which the class of source domain label data ($Y_s$) is {$0$, and $1$}, where the normal scenario is represented by $1$ and the attack scenario is represented by $0$.

\textbf{Target Domain:} The domain has a new type of attacks. The target domain data ($D_t : (X_t,Y_t)$) is the combination of $ {(X_{t1},Y_{t1}), (X_{t2},Y_{t2}), (X_{t3},Y_{t3}), ...... (X_{tn},Y_{tm})}$, in which the class of target domain label data ($Y_t$) is {$0$ and $1$}, where the normal scenario is represented by $1$ and the attack scenario is represented by $0$.

Furthermore, the source domain label ($ Y_s $) and the target domain label ($ Y_t $) contain only ``normal” and ``attack” data, but attackers in the source domain and target domain may be different. Although the source domain label ($ Y_s $) and the target domain label ($ Y_t $)  have the same feature space, their performance in specific features is different. We have used the formula called maximum mean discrepancy (MMD) \cite{MMD} to measure the difference between the source domain and the target domain.

\begin{eqnarray*}
Distance (X_s, X_t) = \Bigg\| \frac{1}{n} \sum_{i=0}^{n} \phi (X_{s_i}) - \frac{1}{m} \sum_{i=0}^{m} \phi (X_{t_i}) \Bigg\|^2 \\
\end{eqnarray*}

According to the dependence of Traditional machine Learning(TML) and DL models, the detection model trained by source domain data ($ D_s $) does not have good detection accuracy when facing target domain data ($ D_t $), and it has been completely confirmed by the subsequent experiment. The TML and DL models need sufficient training data, thus it is difficult to train an efficient APT Detection model model only depending on a small-scale of target domain source data ($ D_t $).

\begin{table}[htp]
\renewcommand{\arraystretch}{1.3}
\caption{Symbols and description}
\label{Table: Symbols and description}
\centering
\begin{tabular}{|>{\rowmac}l|>{\rowmac}c|>{\rowmac}c<{\clearrow}|}
\hline
\setrow{\bfseries}Description & Source (s) & Target (t) \\
\hline
Domain data & $D_s : (X_s,Y_s)$ & $D_t : (X_t,Y_t)$\\
\hline
Domain feature & $ X_s $ & $ X_t $\\
\hline
Domain label &  $ Y_s $ & $ Y_t $ \\
\hline
Number of domain data & n & m \\
\hline
\end{tabular}
\end{table}

Therefore, we transfer the knowledge contained in source domain data ($ D_s $) to the target domain through the proposed DTL-ResNet method and combine the target domain data ($ D_t $) with the same DTL-ResNet method to construct an efficient APT Detection for the target domain to improve the detection accuracy for any heterogeneous IoT ecosystems.

\subsection{DTL ResNet APT Detection Technique}

\begin{figure}[htb!]
    \centering
    \includegraphics[width=\linewidth]{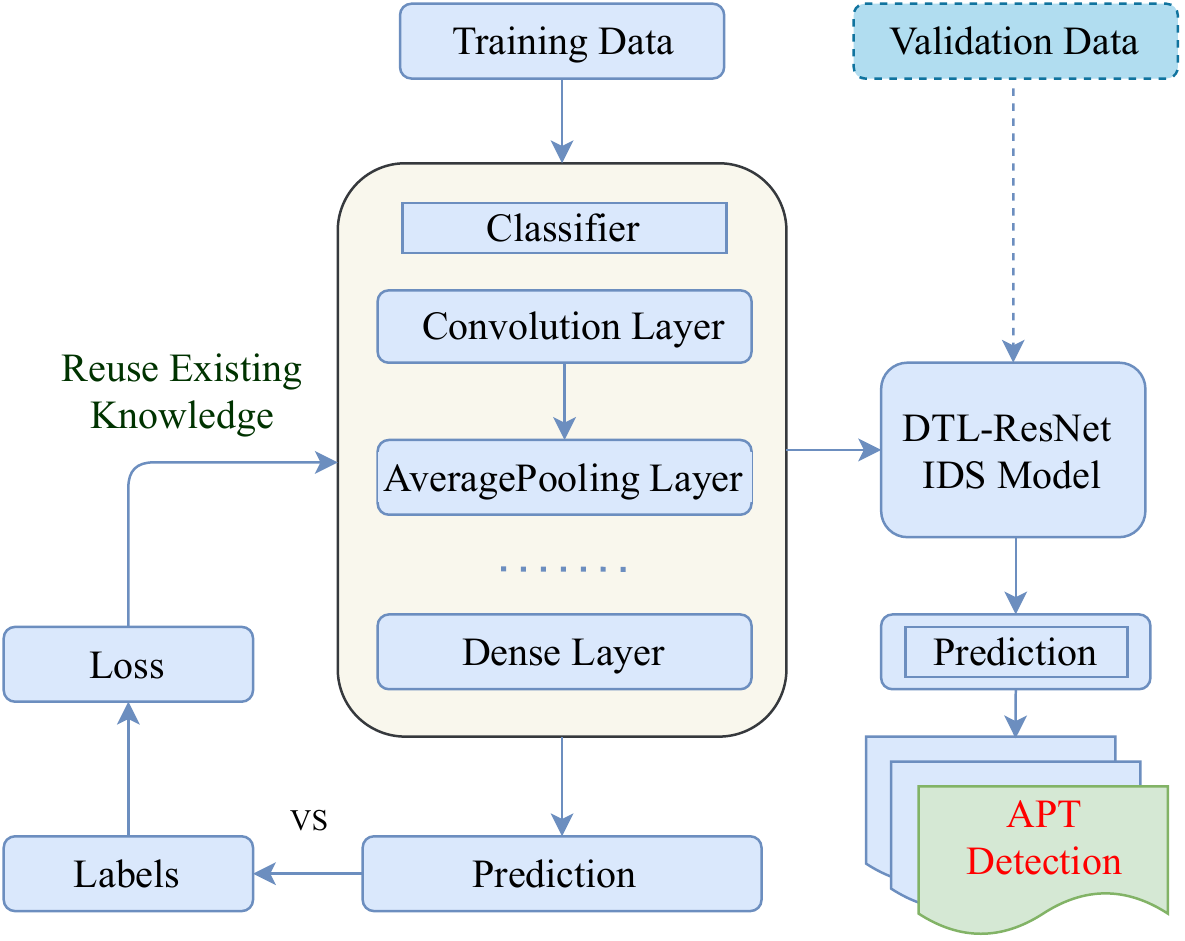}
    \caption{APT Detection flow using DTL ResNet model. The prediction compares with the previous loss and reuse the knowledge. Once model is ready and deployed in the edge-server can detect APT injected to the system.}
    \label{fig:fig_5}
\end{figure}

Figure \ref{fig:fig_5} shows the block diagram of the proposed DTL-ResNet based model for APT Detection, which predominantly includes two parts: the first one is the model training part and the last one is the intrusion detection part. We have applied it to our proposed model after prepossessing the network data. The most significant parameters of this model will be determined through subsequent empirical experiments. The model with an optimal prediction performance on the training set will be selected as the final intrusion detection model for heterogeneous IoT applications.  as for the intrusion detection part, we have trained the DTL-ResNet model by randomly selected training dataset and validated the model by validation dataset. The detection performance of the models under the discrete type of parameters are compared. Finally, the optimal performance model has been selected as the final detection model in the field of heterogeneous IoT applications. 

The network architecture which we have selected for the DTL approach is a one-dimensional Fully Convolutional Neural Network (FCN) called \textit{Conv1D}. Figure \ref{fig:fig_5} shows the architecture of the proposed DTL-ResNet model. The input of the network is the same shape. The one-dimensional convolution layer is used in the first, second and third layers respectively. The first layer is the combination of 128 filters of kernel length 8, the second layer is the combination of 256 filters of kernel length 5, and the third layer is the combination of 128 filters of kernel length 3.  The \textit{Rectified Linear-Unit (ReLU)} activation function is used in the third one-dimensional convolution layer. Each one-dimensional convolution layer is followed by a \textit{BatchNormalization()} operation \cite{Ioffe-46}. The combination of these three convolution layers has a stride equal to 1. The procedure repeat two times for block-2 and block 3 respectively purposing to achieve optimal performance. Each block takes the previous block outputs as inputs for the current block and performs some non-linearity's to transform it into a multivariate series whose dimensions are defined by the number of filters in each layer. The fourth layer is the combination of a \textit{GlobalAveragePooling1D()} operation which takes the input of the third block and averages each series. This operation reduces drastically the number of parameters in a deep model while enabling the use of a class activation map \cite{Zhou-47} which allows an interpretation of the learned features \cite{Ioffe-46}.  The output of the gap layer is then fed to a \textit{softmax} classification layer whose number of neurons is equal to the number of classes in the dataset.  Other hyper-parameters and data corelation are skipped purposing brevity of the paper. This enabled us to identify the effect of deep transfer learning in the field of APT detection. However, once APT detection successfully run in the edge-server, the detection status is kept encrypted purposing to send it during the data transfer over the consortium blockchain (CBC) network. Suppose, a detection status is $T_s$ and data transaction is $T_x$. This will be recorded in the blockchain ledger (BCL). As mentioned earlier the CBC works as KGD which establishes the communication between edge-devices and blcockhain peers. Though, existing consortium blockchain i.e., hyperledger fabric (HLF) \cite{GDPR2019} works in cooperation with the membership service provider (MSP) which is actually a PKI-driven certificate authority (CA), the proposed technique replaces the need of CA need by facilitating a novel certificateless technique using Elliptci curve powered multisiganture (MS), i.e. BLS/Schnorr variant etc. The following subsection explains how it fullfills that requirements.

\subsection{Blockchain based Certificateless Authentication}

Identity-based encryption (IBE) encounters crucial key escrow issues while it emerges with the legacy of Public Key Infrastructure (PKI). As discussed in the previous section certificate-less cryptography (CLC) solved it by introducing the partial secret $(ps)$ concept, derived from a master secret $(ms)$ that keeps the private keys apart from blockchain consortium ($KGD$) as already mentioned in Figure \ref{fig:fig_CL}. The $(ps)$ depends on the device identity, which furthermore ensures the mutual dependency instead of sole reliance on trusted TTP, i.e., certificate authority (CA). In Industry 4.0 use case, once a device receives the partial secret $(ps)$, it starts generating public-private key pairs ($Pk, Sk$) using its identities ($Ids$). As depicted by the 4 steps ( $a:$ to $d:$) interaction between $A$ and $B$ of Figure 4,  the entire key generation tasks can be divided into the following five (05) consequent processes.

\subsubsection*{setup$(1^\lambda)\rightarrow (y, ms)$} It takes a system's security parameter $\lambda$ and returns the system parameter $y$ and master secret $(ms)$. The algorithm associated with this procedure runs at ($KGD$). For example, for the customized Schnorr multisignature (MS), $y$ includes the prime numbers$(p,q)$, group generators $(g)$ i.e. primitive roots etc. It finalizes the after confirming each peer's own private keys such as $(x_1,x_2,x_3$, master secret $(ms)$ and public keys such as $(p,q,g,y$ as mentioned in the earlier key generation subsection.

\subsubsection*{$genPS(y, id_j, ms) \rightarrow (ps_j)$} As illustrated by Algorithm 1 the algorithm takes inputs of system parameter $y$, $j$'th number of identities $(id_j$ that interested to join at a certain time $t$, along with the previously created master secret $(ms)$. Here the device identity, $id_j \in \{0,1\}^*$, and $ms$ outputs the $j$'th number of partial keys $(ps_j)$ in response. In line with that, it takes the $(id_j)$ and finds device secret value $x_j \leftarrow (y, id_j)$.

\subsubsection*{$multiSig(x_i, ps_j)\rightarrow (R,S)$} At this stage, Blockchain peers cosign the partial secret $(ps_j)$ before sending it to the IIoT sensors through edge platform. Initially, they compute the shareable parameters $R_i\leftarrow g^r_i$ after choosing own secret random number $r_i$. Then aggregates $R \leftarrow \prod_{i}^{n} R_{i}\: mod \: p$. Besides, it computes, $c\rightarrow (T\parallel Y \parallel R \parallel ps_j)$,  individual signature, $s_i \leftarrow (r_i + cx_i) \ mod \ q$ and lastly aggregate $S\leftarrow  \sum_{i=i}^{n} s_{i}\: mod \: p$. Therefore, the signature outcomes as if $\sigma \leftarrow (R,S)$.

\subsubsection*{$genSk(y, ps_j, x_j) \rightarrow SK_j$} At the beginning of this step, the Edge devices should know the system security parameters, $Y$ which includes the prime numbers $(p,q)$ group generating primitive roots $g$ and calculated public key variant $y$.Then, edge gateway on behalf of device or the each IIoT device itself verifies if the partial keys received are valid or not simply by justifying the conditions such as $g^S \ mod \ p = R \times y^c \ mod \ p$. If the multi-signed partial keys $(ps_j)$ are valid, the particular IIoT device run its algorithm to produce its own private keys $(sk_j)$.

\subsubsection*{$genPk(y, x_j)\rightarrow pk_j$} To generate the public keys $(pk_j)$ for the IIoT devices, it takes system parameters along with the previously generated device secret $x_j$. Please note that, though it looks similar but the public generation for the devices are not the same process as the key generation of the Blockchain peers as discussed earlier.

The above procedures should work in line with some base method such $encrypt()$, $decrypt()$ and $verify()$ along the associated algorithms that could be simplified as below. 

\begin{enumerate}[label=\alph*)]
    \item $enc(y, ps\in m, id, pk)\rightarrow (c\in  \mathbb{c}\vee \bot)$: The base $encryption$ method takes the system parameters $y$, partial secrets $(ps)$ as message $(m)$ and produces the desired ciper-text $(c)$ within the designated cipher-text space.
    \item $dec(c\in \mathbb{c}, sk\rightarrow ps\in m \vee \bot)$:The base $decryption$ method and its associated algorithm takes the cipher-text and produces the partial-secret $(ps)$ within the designated message space.
    \item $ver(y, \sigma, id, ps)\rightarrow true (1) \vee false(0) \vee \bot$:The verification algorithm in the edge side, takes the system security parameters $(y)$, the multi-signed signature $(\sigma)$, device identities $(id)$ and the partial secrets $(ps)$, hence verifies either the multi-signed partial secrets are valid $(true \lor 1)$ or invalid $(invalid \lor 0)$. 
\end{enumerate}

\begin{algorithm}

% functions

\SetKwFunction{setup}{setup}
\SetKwFunction{keyGen}{keyGen}
\SetKwFunction{genSk}{genSk}
\SetKwFunction{requestSend}{requestSend}
\SetKwFunction{genPS}{genPS}
\SetKwFunction{responseReceived}{responseReceived}
\SetKwFunction{multiSig}{multiSig}
\SetKwFunction{genPS}{genPS}
\SetKwFunction{verify}{verify}
\SetKwFunction{genPk}{genPk}

% input/ouput names
\SetKwInOut{Input}{Input}
\SetKwInOut{Output}{Output}

% caption
\caption{Certificateless device key generation \label{alg:singlepm}}

\Input{%
		\xvbox{4mm}{$\xvar{id}_j$} -- identities of the $j'th$ number of IIoT devices  \\
		\xvbox{2mm}{$\xvar{y}$} -- system parameters \tcc*{prime numbers}
	  }
\Output{%
		\xvbox{12mm}{$\xvar{(pk, sk)}$} -- public and private key pairs
		} 

  \BlankLine % blank line for spacing
  
  % start of the pseudocode
  \xvbox{12mm}{$\xvar{\setup}(1^\lambda)$} $\rightarrow (y)$ \tcc*{sys param init}
  
  \For{$\xvar{id}  \leftarrow$ $\xvar{id}_j$ }{
  
  \xvbox{2mm}{$\xvar{porocedure \keyGen(y, id):}$}  \tcc*{key gen}

  \xvbox{2mm}{$\xvar{X}_{ j }$} $\leftarrow$ \genSk(y,$\;id_j$) \tcc*{gensec keys}
  
  \xvbox{2mm}{$\xvar{\requestSend(id}_j)$}  \tcc*{request to join}
  
  \xvbox{4mm}{$\xvar{ps}_{ j }$} $\;\leftarrow$ \genPS($id_j$) \tcc*{KGDs gen partial sec}
  
  \xvbox{12mm}{$\xvar{\multiSig}(s_n, id_j, ps_j)$}  \tcc*{multi-sig}
    
  \xvbox{2mm}{$\xvar{\responseReceived(id}_j)$}  \tcc*{get  $ps$}
  
  \xvbox{14mm}{$\xvar{V\;\;[0,1,\xvar{$\bot$} ]}\leftarrow\;$$\xvar{\verify()}$}   \tcc*{verify sig}
    
  \If{$\xvar{V}\leftarrow$ $\xvar{1}$ }{

    \xvbox{4mm}{$\xvar{sk}_j$} $\;\leftarrow$ \genSk(Y,$\;id_j$,$\;x_j$) \tcc*{set pri key}
    
    \xvbox{4mm}{$\xvar{pk}_j$} $\;\leftarrow$ \genPk(y, $\;x_j$) \tcc*{sets pub key}
  } % end for j	
  }

\end{algorithm}

\subsection{Device Registration and Verification}
$KGD$ plays an indispensable rule for security of the edge sensors. The sensor devices need to get registered with the $KGD$ consortium. Before posting the transaction ($t_x$), interested sensors obtain public-private key pairs upon the completion of the registration process. 
\subsubsection{Registering devices} As discussed earlier, multi-party industry 4.0 stakeholders agrees to build the $KGD$ cooperatively. Suppose, the owner, buyer and insurer along with other stakeholders cooperatively form the BC network that represents the $KGD$ peers. They peers broadcasts the system parameters $(y)$ to  all sensors. $KGD$ peers keep their individual signer’s secret such as $s_1$, $s_2$...$s_n$ With the help of Edge computation capacity or its own ability, interested  device creates their own secret value $x_1, x_2, x_3...x_j$ generates respective public keys using $x_j$ and the system parameter $y$, where $j$ is the number of interested devices at particular time $t$ and $n$ is the number of co-signing Blockchain peers.  Sensor devices will contact the KGD with their identities $id_1, id_2,...id_j$. Upon receiving the request, $KGD$ will generate a partial private seccret $ps_1, ps_2...ps_i$ for all requested devices and will cosign co-sign $id_i$ and $ps_i$ using co-signers private key $s_n$. KGD sends the signed message back to the edge. The sensor device itself or edge node (e.g. Azure IoT edge or Dell Gateway) will verify if the message comes from the KGD, and if yes, it will generate private - private key pairs $(pk_1, pk_2,...,pk_i , sk_1, sk_2,...,sk_i)$  using $(ps_i,x_i,Y$. Note that only each industrial IoT device will be able to create the private key because it is the only entity who knows his private secrets $x_j$. \textbf{Alg}. \ref{alg:singlepm} illustrates the step by process with the necessary explanations.

\begin{figure}
    \centering
    \includegraphics[width=\linewidth]{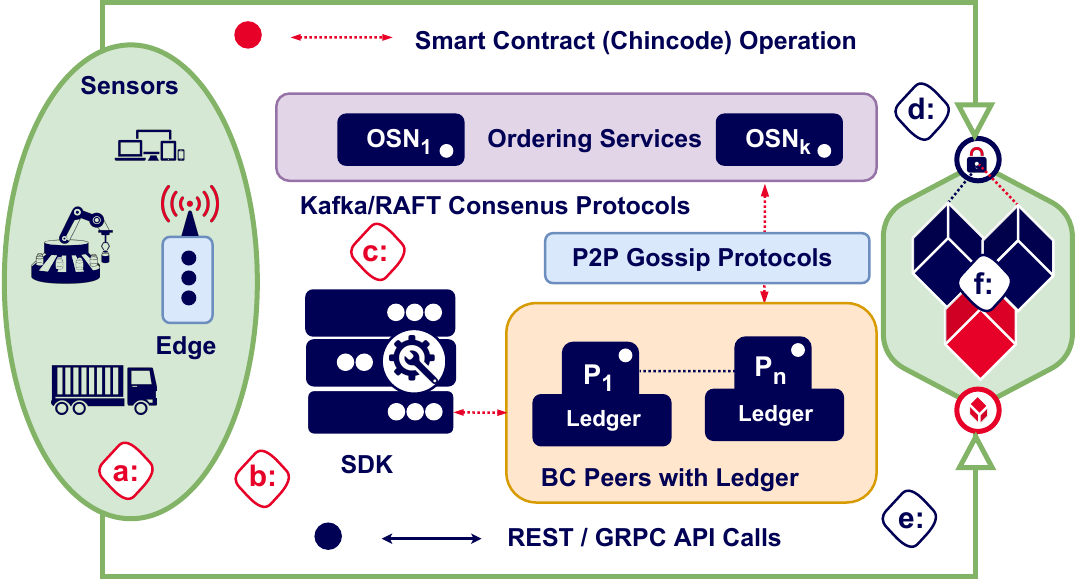}
    \caption{Consortium blockchain (hyperledger fabric) data transaction ($T_x$), verification and recording  flow. It has 6 steps, i.e., \textit{a :} Proposing Tx ( registrar  \& submit Tx), \textit{b :} Endorsing Tx (run environment \& endorse), \textit{c :} Verifying Tx (validate propsed Tx ), \textit{d :} Aggregating Tx (OSN articulate), \textit{e :} Committing Tx (broadcast to all), \textit{f :} Storing Tx ( to ledger \& DHT )}
    \label{fig:fig_66}
\end{figure}

\subsection{Data Protection and Storing}

Once edge sensors are successfully registered to the $KGD$ upon the certificate-less cryptography and multi-signature based authentication, the sensor devices proceed further to send and store data as illustrated by Figure \ref{fig:fig_66}. The transaction includes the identity of the IIoT devices along with the action and timestamp at the time $(T)$ of action ($ACT$). There can be different type of actions such as $store$ data at a specific DHT address ( $ADS$), $update$ previously inserted data or $access$ permission of the particular data. To verify a transaction $T_x = (ID_j,T, ACT)$, the blockchain peers have to meet two conditions: \textit{i)} Either the public key ($PK_j$) obtained associates with the identity ($ID_j$), \textit{ii}) Can the signed transaction ($T_x$) be verified. Figure 5 illustrates the following steps from \textit{1} to \textit{6}.

\textit{\textbf{a) Propose}}: Client Edge sensors initiate the process by registering the devices to the Blockchain. It constructs the encrypted transaction proposal ($t_x$) using $(s_k)$  and invoke the smart contract and SDK.
    
\textit{\textbf{b) Endorse}}: SDK requests for endorsement, and BC peer verifies $t_x$ after authenticating the $id$.
    
\textit{\textbf{c) Verify}}: The verification requires meeting the policy, i.e., business logic. The SC takes a $t_x$ as input and returns a multi-signed $0$  or $1$  in response to the SDK apps.
%The cosigning operations are achieved through the process discussed earlier. In certain cases, the action process diverges for either query request to retrieve data from or data transaction to store into.}%
$T_x$ is determined as query function using APIs (i.e OAuth 2.0 REST API). In either case, the SDK apps proceed the $t_x$ with the required operations such as $create, retrieve, update$, and $delete$ with the endorsement. 
    
\textit{\textbf{d) Aggregate}}: The SDK apps aggregates all consents into single transaction and disseminates those to the Ordering Service Node (OSN). The OSN works on the consensus protocols, i.e., Practical Byzantine Fault Tolerance (PBFT) within Apache Kafka platform.
    
\textit{\textbf{e) Commit}}: The $t_x$ then relayed to the OSN, associated channel peers confirm each $t_x$ of the block by specific smart-contract and checking through concurrency control cersion (CCV). In case any transaction misses the process is identified with an invalid status inside that block. Hence a fresh block is committed to the blockchain ledger.

\begin{algorithm}

% functions
\SetKwFunction{create}{create}
\SetKwFunction{signTx}{signTx}
\SetKwFunction{castTx}{castTx}
\SetKwFunction{verID}{verID}
\SetKwFunction{verTx}{verTx}
\SetKwFunction{storeDHT}{storeDHT}

% input/ouput names
\SetKwInOut{Input}{Input}
\SetKwInOut{Output}{Output}

% caption
% \addto\captionsenglish{\renewcommand{\algorithmcfname{Alg.}}

\caption{Secure data transfer and store by BC SC\label{alg:veri-and-store}}

\Input{%
		\xvbox{4mm}{$\xvar{T}_x$} -- IIoT data transactions  \\
		\xvbox{2mm}{$\xvar{L}$} -- access control lists \\
		\xvbox{2mm}{$\xvar{}$$\sigma$} -- signaturues of the $T_x$ \\
		\xvbox{4mm}{$\xvar{ID}_j$} -- identities of the $j'th$ number of IIoT devices  \\
		\xvbox{2mm}{$\xvar{Y}$} -- system parameters \tcc*{prime numbers, primitive roots etc}
	  }
\Output{%
		\xvbox{17mm}{$\xvar{($V_{I_d}$,$V_{T_x}$, S)}$} -- set \& return verification and storing flag \textit{true}
		} 

  \BlankLine % blank line for spacing
 
  \xvbox{2mm}{$\xvar{\create \; :=  (ID,\;L,\;Tx,\;$\sigma$,\;ADS)}$}  \tcc*{creates $Tx$ }
  
  \xvbox{2mm}{$\xvar{\signTx(Tx},Sk)$}  \tcc*{sign $Tx$}
  
  \xvbox{2mm}{$\xvar{\castTx(Tx},\sigma)$}  \tcc*{broadcasts $Tx$}
  
  \For{$\xvar{Tx}  \leftarrow$ $\xvar{Tx}_i$ \tcc*{for all $n\times T_x$} }{
  
  \xvbox{2mm}{$\xvar{$V_1 \leftarrow$ \verID(ID},Pk,Y)$}  \tcc*{verify ($ID$)}
  
  \xvbox{2mm}{$\xvar{$V_2 \leftarrow$ \verTx(Tx,ID},Pk,\sigma)$}  \tcc*{verifies ($Tx$)}

  \If{$\xvar{(V}_1\;||\;V_2)$ }{

    \xvbox{2mm}{$\xvar{S $\leftarrow$ \storeDHT(Tx, ID):}$}  \tcc*{store $Tx$}
  }
  }

\end{algorithm}

\textit{\textbf{f) Store}}: The Gossip protocol of the OSN broadcasts ledger update across the BC network. Thus pointer is memorized in the ledger, and data-address is securely stored on the offline or online DHT data repository upon IPFS or HL CouchDB. Here, the signature algorithm can be represented as a triple /4-tuple of probabilistic polynoimial-time algorithms ($G,S,V$) or ($G, K, E,D$) that includes generation ($G$), $signing(S)$, verification($V$), Key-distribution($K$), Encryption($E$) and Decryption($D$) respectively.  Besides, the identities $ID_j$, here the devices require the Access Control List (ACL) before Transaction ($Tx$) creation and signing ($\sigma$). The industry 4.0 devices along with the Edge Gateway are solely responsible to create the ACl list ($L$) in addition to signature ($\sigma$) generation and transaction($T_x$) publishing. However, the same $L$ will be required later to access data. The algorithm as shown in \textbf{Alg}.\ref{alg:veri-and-store}, outcomes three different flags ($V_1,V_2,S$) set after successful execution. If the identities belong to the derived public keys, $V_1:=true$, while the certificate-less signature meets the condition as discussed earlier, ($V_2:=true$). The Blockchain peers do the transaction ($T_x$) verification in response to the reception. Interchangeable verification procedure works in case of data accessing. Similarly, after data  is written to the DHT, the third flag gets set,($S:=true$). Lastlt a new block is added to the blockchain and subsequently the ledger gets updated including the $T_x$ Pointer ($Tp$).

\section{Dataset and Model Training}
There are two fundamental steps that applied during the data preparation process based on the combined dataset applied \cite{TonIoTDS}\cite{BotIotDS}. Some features such as - date, time, and timestamp have been omitted from feature vectors as they may cause to overfit the training data. Furthermore, for some DL and DTL models, the input data shape has been reshaped into three dimensions to feed the models by applying \textit{numpy.reshape} with \textit{swapaxes} and \textit{concatenate} methods. As this dataset originates from multiple heterogeneous sources. So, it is an essential step to combined all the IoT sensors' data by \textit{redundancy} and \textit{correlation} analysis which has been evaluated using the \textit{Pearson’s product-moment coefficient} equation \cite{Chee-57}.

\begin{equation*}
  r =
  \frac{ \sum_{i=1}^{n}(x_i-\bar{x})(y_i-\bar{y}) }{%
        \sqrt{\sum_{i=1}^{n}(x_i-\bar{x})^2}\sqrt{\sum_{i=1}^{n}(y_i-\bar{y})^2}}
\end{equation*}
where n is the number of tuples, x\textsubscript{i} and y\textsubscript{i} are the respective values in tuple i, $\bar{x}$ and $\bar{y}$ are the respective mean values of $x$ and $y$.

As datasets originate from different heterogeneous sources, it is an essential step to combine all the IoT sensors' data by \textit{redundancy} and \textit{correlation} analysis. This analysis has measured how strongly one feature, i.e., \textit{door\_state} implies the other, i.e., \textit{light\_status}. Table II shows the respective correlation analysis where light status (LS), door state (DS), Smartphone signal (PS), temperature condition (TC), pressure (PS), current temperature (CT), humidity (HY), temperature (TE) are in the rows. Similarly thermostat status (TS), motion status (MS), longitude (LG), latitude (LT), fridge temperature (FT), temperature (TE), humidity (HY), current temperature (CT) are in the column, respectively. We have performed these strategies to scale the selected feature values within a range between [0.0,1.0] using a technique called \textit{minimum-maximum normalization}\cite{Aminanto-58}. 

\begin{equation*}
      N_{ormalized}V_{alue} = \frac{(X-X_{min})} {(X_{max}- X_{min})} 
\end{equation*}

Where, X is an original value and X\textsubscript{max} and X\textsubscript{min} is the maximum and minimum values of the feature, respectively.

\begin{table}[ht]
\caption{Correlation matrix  (with color intensity)}
\centering
\begin{tabular}{|c|c|c|c|c|c|c|c||c|}
\hline
\hline
  \textbf{LS} & \textbf{DS} & \textbf{PS} & \textbf{TC} & \textbf{PR} & \textbf{CT} & \textbf{HY} & \textbf{TE} & \\ \hline \hline
  0 & 0 & 0 & 0 & \textcolor{cor-very-weak}{0.12} & 0 & 0.01 & \textcolor{cor-weak}{0.15} & \textbf{TS} \\ \hline 
  0 & 0 & 0 & 0 & 0.02 & 0 & 0 & 0.01 & \textbf{MS} \\ \hline
  0 & 0 & 0 & 0 & 0.02 & 0 & 0 & 0.03 & \textbf{LG} \\ \hline
  0 & 0 & 0 & 0 & 0.02 & 0 & 0 & 0.03 & \textbf{LT} \\ \hline
  0 & 0 & 0 & 0 & \textcolor{cor-very-strong}{0.57} & \textcolor{cor-very-weak}{0.16} & \textcolor{cor-very-weak}{0.35} & \textcolor{cor-very-strong}{0.60} & \textbf{FT} \\ \hline
  0.01 & \textcolor{cor-very-weak}{0.16} & 0.01 & 0.01 & 0 & \textcolor{cor-very-strong}{0.58} & 0 &  & \textbf{TE} \\ \hline
  0 & 0.01 & 0 & 0 & 0.04 & \textcolor{cor-very-weak}{0.33} &   &  &  \textbf{HY} \\ \hline
    0 & 0 & 0 & 0 & \textcolor{cor-very-strong}{0.55} &  &  &  & \textbf{CT} \\ \hline
  0 & 0.12 & 0 & 0.01 &  & &  &  & \textbf{PS} \\ \hline
  
\end{tabular}
\label{tab:correlations-matrix}
\end{table}

\subsection{Training the DTL ResNet Model}
First of all, we have divided the combined dataset into the training data set (80\%) and the validation data set (20\%) by using the \textit{train\_test\_split} method of the \textit{scikit\_learn} library. To avoid the over-fitting problem, this splitting ratio has been considered as the best ratio between the training and the validation dataset \cite{Guyon-60}. We have used the value of the \textit{random\_state} parameter as true (1), which decided the splitting of data into the training and the validation set randomly. The k-fold \textit{cross\_validation} has been used for parameter tuning.

\section{APT Detection Evaluation}
First of all, we consider the quantitative performance of DTL algorithms. Table \ref{Table: DTL performance comparison metrics} shows the quantitative performance summary of the DTL algorithms, where the proposed ResNet model shows an optimal performance compared to the other DTL algorithms with an accuracy score of 0.87, precision score of 0.88, recall score of 0.86, f1-score of 0.86, and ROC AUC score of 0.83. In this model, we used three hidden-layers where \textit{relu} is the hidden layer activation function. Also, \textit{softmax} is used as a network output activation function, and ``categorical\texttt{\_}crossentropy" is used as a loss function along with \textit{adam} optimizer.

\begin{figure}
    \centering
    \includegraphics[width=\linewidth]{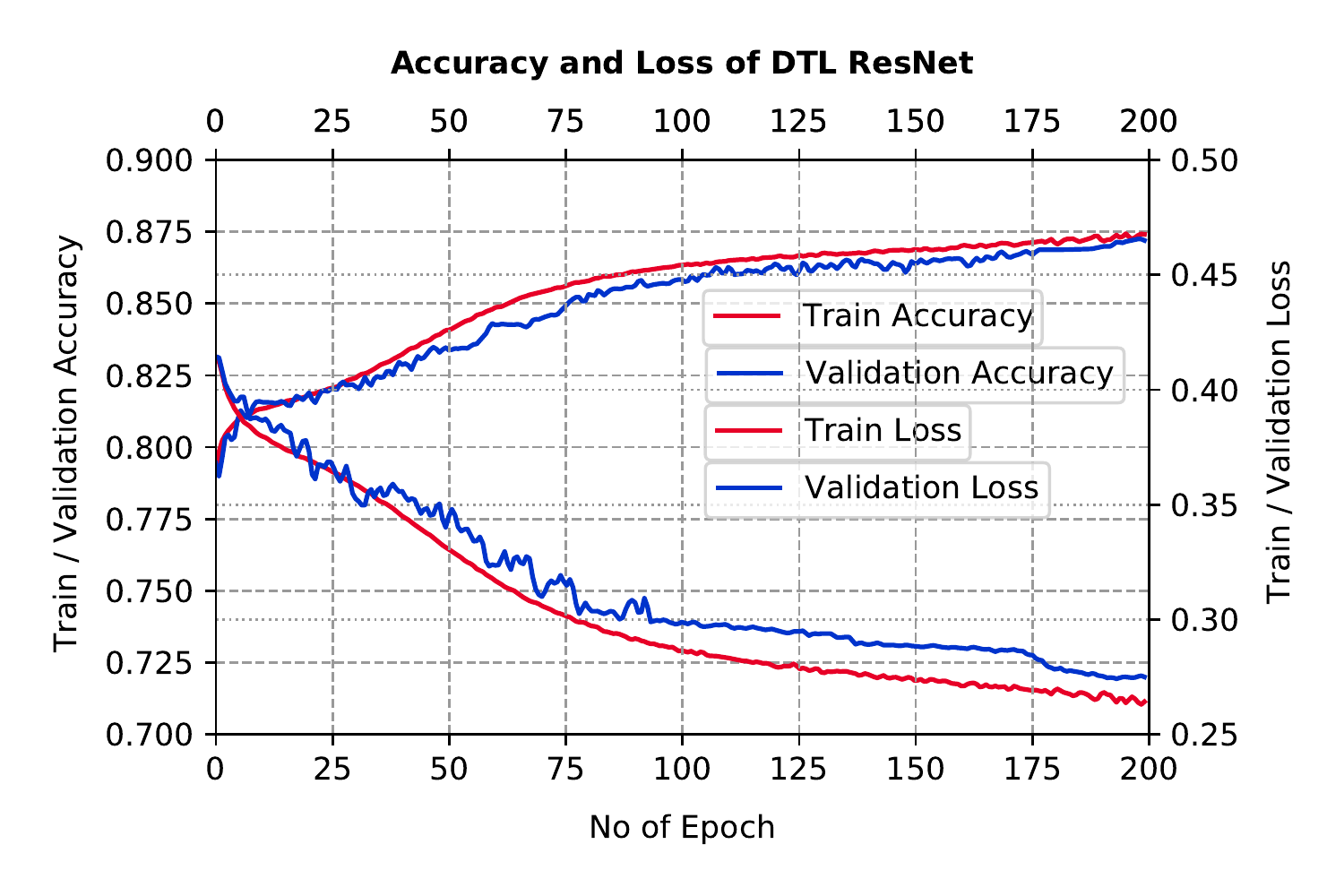}
    \caption{The training and validation accuracy of the ResNet model used to detect Advanced Persistent Threat (APT). The experiment performed within an edge complaint setup.}
    \label{fig:fig_67}
\end{figure}

\begin{table}[htp]
\renewcommand{\arraystretch}{1.3}
\caption{DTL performance comparison metrics}
\label{Table: DTL performance comparison metrics}
\centering
\begin{tabular}{>{\rowmac}c>{\rowmac}c>{\rowmac}c>{\rowmac}c>{\rowmac}c>{\rowmac}c<{\clearrow}}
\hline
\setrow{\bfseries}Algorithm & Accuracy & Precision & Recall & F1Score & ROC AUC\\
\hline
FCN & 0.84 & 0.85 & 0.84 & 0.83 & 0.81\\
\hline
LeNet & 0.80 & 0.82 & 0.80 & 0.79 &0.76\\
\hline
IncepNet & 0.80 & 0.86 & 0.80 & 0.81 & 0.73\\
\hline
MCDCNN & 0.80 & 0.83 & 0.80 & 0.79 & 0.76\\
\hline
CNN & 0.81 & 0.83 & 0.81 & 0.80 & 0.76\\
\hline
LSTM & 0.85 & 0.84 & 0.77 & 0.77 & 0.83\\
\hline
MLP & 0.73 & 0.74 & 0.78 & 0.77 & 0.74\\
\hline
\setrow{\bfseries}ResNet & 0.87 & 0.88 & 0.86 & 0.86 & 0.83\\
\hline
\end{tabular}
\end{table}
Figure \ref{fig:fig_67} shows the accuracy score of every single epoch for both training and validation phases on the above-mentioned eight discrete DTL models. Considering epoch number 170 to 200 for both phases, the flattening characteristics of the curve and accuracy are not increasing literally, we have considered 200 epochs for our analysis. The proposed ResNet model (P-ResNet) shows the highest accuracy score in both the training and validation phase. Whereas, MLP model shows the lowest accuracy score during any settings of the epoch number within 1 to 200. In detail, according to Figure \ref{fig:fig_67}, the accuracy of the MLP model starts around 0.67 for the training phase and 0.63 for the validation phase in the epoch number 10. But it increase dramatically around 0.78 in epoch number 65 and 0.74 in epoch number 140 for the training and validation phase, respectively.

However, for the training phase, the accuracy score (0.78) remains stable in epoch number 66 to 200. On the other hand, for the validation phase, the accuracy score (0.74) remains stable between epoch numbers 141 and 200. The training and validation accuracy of CNN, IncepNet, LeNet, and MCDCNN models remain steady between the epoch number 1 to 200 as shown in figure \ref{fig:fig_67}. The accuracy of LSTM and FCN models starts with a score of 0.82 at the beginning. But this score rises gradually with the increase if the epoch number and reaches to approximately 0.86 when the epoch number is 160 and then remains stable between epoch number 161 to 200 for both of the phases. The remarkable point is that the behavior of the training phase almost similar to the validation phase. Figure \ref{fig:fig_67} shows the trend of the accuracy score of both phases for a better understanding of our proposed model. The accuracy of the proposed model jumps rapidly in epoch number 60 and it reaches a peak of point close to 0.87 at epoch number 169. However, According to Figure \ref{fig:fig_67}, which remains almost stable up to the early stopping checkpoint with an accuracy of 0.87.

\section{Blockchain Performance Evaluation}
\subsection{Platform Setup}
The proposed security approach was deployed inside the Caliper evaluation toolkit for IBM Hyperledger Fabric (HLF) v1.4.1. It helps measuring a particular blockchain deployment with a set of previously defined enterprise use cases. IBM discloses that no general tool provides performance evaluation benchmark for Blockchain while releasing initial version of HLF Caliper\cite{GDPR2019}. The integrated use-cases were customized to overlap the industry 4.0 edge requirements for generating data. However, the latest version of the NodeJS Package Manager (NPM 8.0.1), docker, and  curl  were installed to set up the run-time environment inside the Ubuntu 18.04 LTS with 16 GB of memory where \textit{python2} , \textit{make}, \textit{g++} and git ensure additional SDK supports. A typical configuration for the permissioned blockchain has programs called \textit{Test Harness} that include client generation and observation and the deployed blockchain System Under Test (\textit{SUT}) and the RESTful SDK \cite{GDPR2019}.

\subsection{Blockchain Deployment} The RESTful Software Development Kits (SDK) interfaces among the required components setup. There are four (04) steps required to evaluate the performance benchmark, such as \textit{i}) Starting a local Verdaccio-server for package publishing, \textit{ii}) The connecting the repository to the server, \textit{iii}) Installation and binding the CLI from the server side and \textit{iv}) Hence, running the integration benchmark. The associated ledger works with the initial \textit{config.yml} file on command line interface (CLI).%\footnote{HLF configuration files https://github.com/rahmanziaur/IIoTConsortiumBC}% 
After the initial configuration, the system was configured for performance benchmark with the  tasks, such as - \textit{a}) invoke policy checking functions (READ) and WRITE $Tx$ into the ledger, \textit{b}) Setup multiple test-cases for about 2 to 35 number of peers representing industry stakeholders and cosigners, \textit{c}) Allocating workloads from 100 Tx/sec to 1500 Tx/sec among those peers representing the Industry 4.0 edge data population. However, the future scope of this work includes increasing the workload to best suit the higher Industry 4.0 CPS standard.

\begin{figure}%[htb!]
    \centering
    \includegraphics[width=\linewidth]{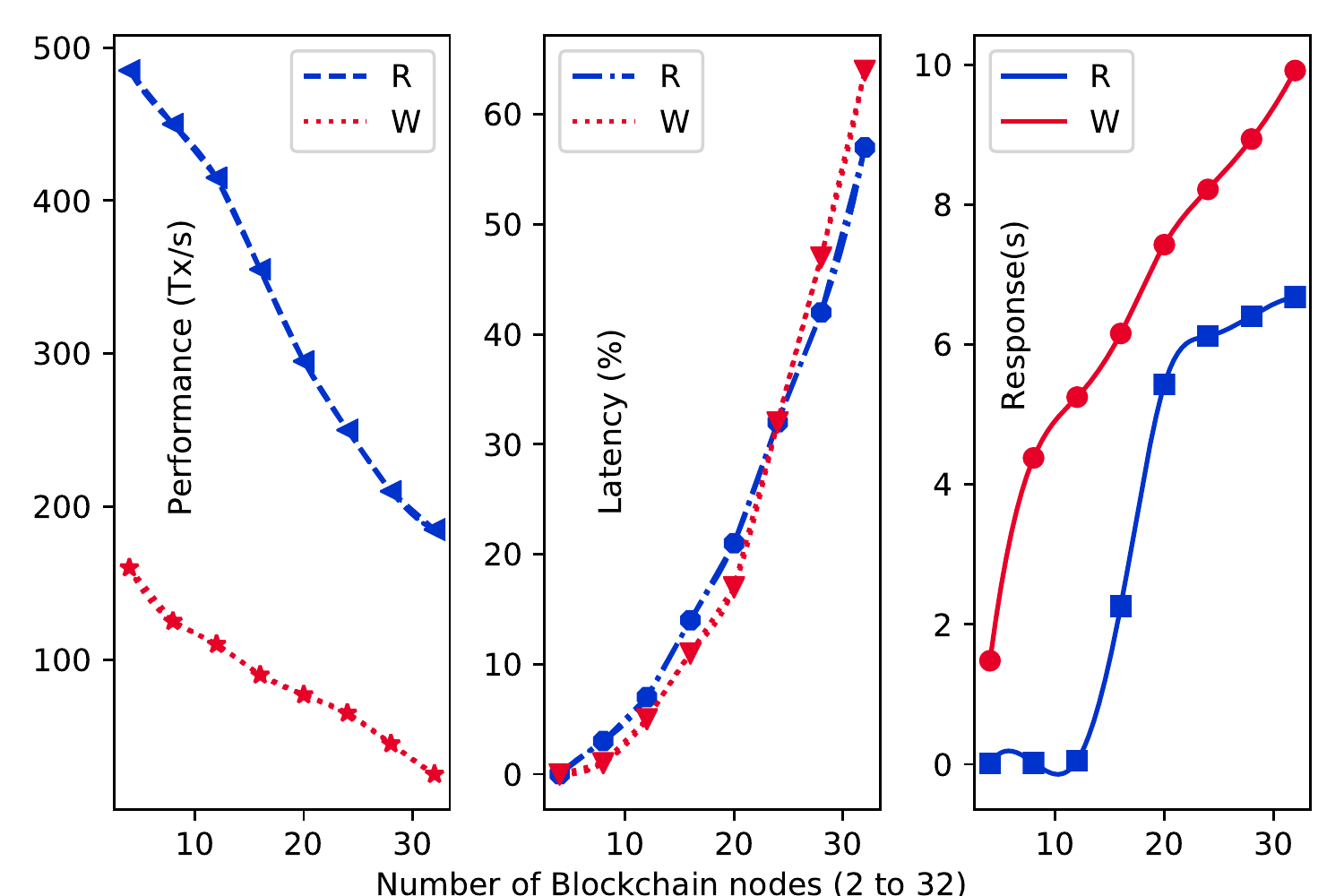}
    \caption{System performance based on Blockchain transaction processing rate. Here (a) Performance (b) Latency (c) Response time for the use case with 2 to 32 nodes.}
    \label{fig:fig_78}
\end{figure}

\subsection{Performance Analysis}

The Caliper benchmarking results illustrates the deployed project performance based on four measurement metrics success rate ($\rho$), latency ($\Delta$t and $L$), throughput ($P$), and resource consumption ($W$) for different test cases. Fig \ref{fig:fig_78} shows the throughput, success rate and delay, respectively.  The associated test-case was run under different number of workload ($W$) ranging from 100 to 1000 workloads.  The HLF network occupies two (02) chaincodes, four (04) peer nodes, and three (03) OSNs running on Apache Kafka for Practical Byzantine Fault Tolerance (PBFT) consensus algorithm. As seen in Figure \ref{fig:fig_78}(a), the WRITE has 185 at about 200 workload ($W$) with the maximum success rate of 93\% and an average delay of 5 seconds. On the other hand, Read operation seems to have a maximum of 470 throughputs on a similar success rate at the maximum workload. The usual delay appears to be almost half of the $W$ delay as $W$ has to incorporate OSNs on Kafka. 

\begin{figure}
    \centering
    \includegraphics[width=\linewidth]{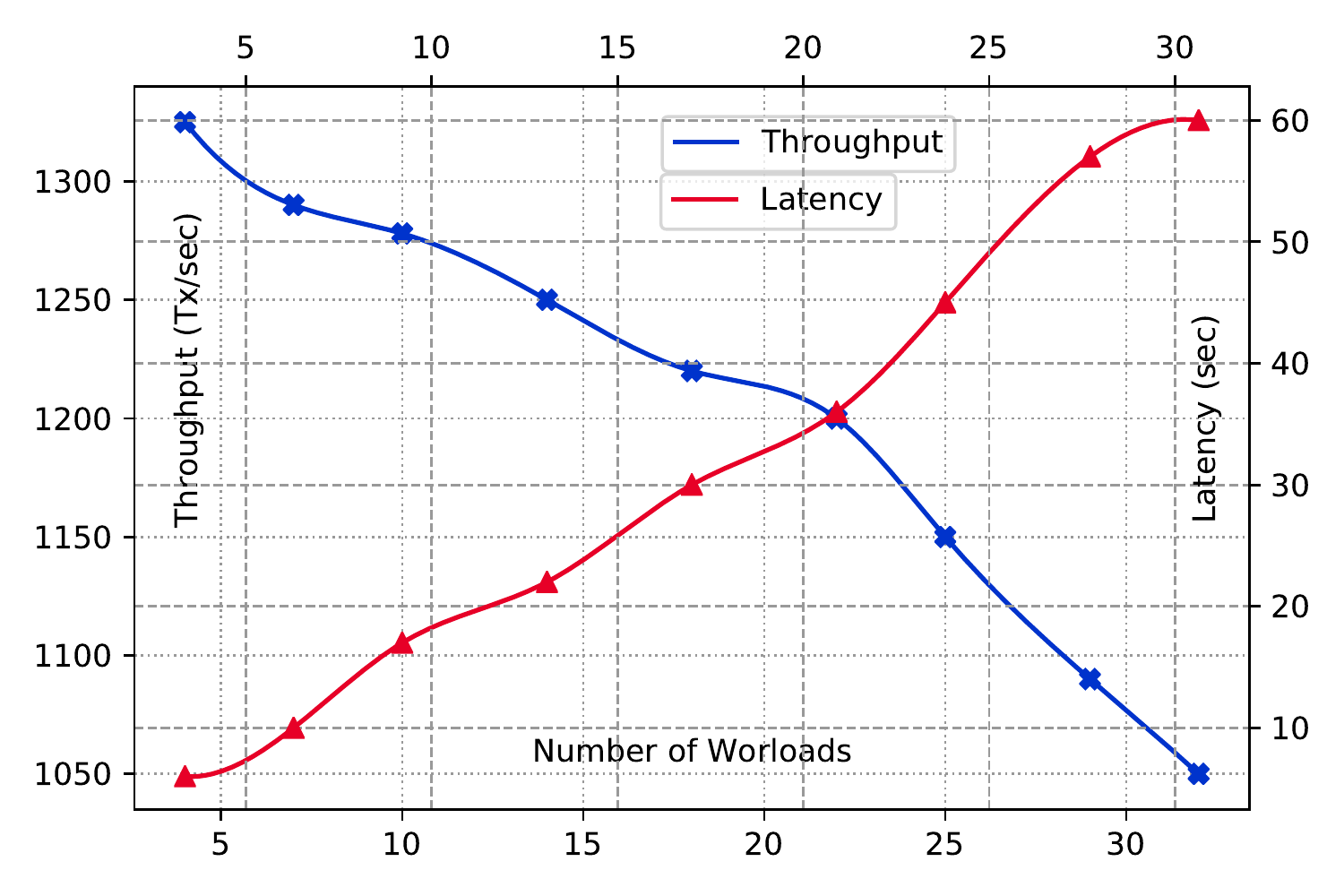}
    \caption{Proposed system scalability. It shows performance based on an average \textit{throughput} vs \textit{latency}}
    \label{fig:fig_89}
\end{figure}

The benchmark evaluation explicitly illustrates that the setup configured has lower performance for higher number workload ($W$) though the theoretically solution proves the consortium Blockchain has significant adaptability for higher number of nodes. As investigated the deep inside,the local workload processing bottleneck affects throughput and latency. Hyperledger $T_x$ flow works demands enough responses against the submitted $T_x$ proposals, in case the responses are queued due to network overhead, bandwidth or processing loads consequences the latency raising. On top of that, the general purpose workstation configuration slower the evaluation for higher workloads. Here, Figure \ref{fig:fig_89} portrays the relation between performance and scalability based on the previously executed Read, Write Operations. To avoid further complexity, OSN and peer configuration left resembling to initial setup. However, two test cases run for 300 and 500 workload. As depicted by Figure \ref{fig:fig_89}, the HLF platform setup has lower scalability. For the first test-case (300 workload), the throughput and latency respectively reaches 150 tps and 64. However, for the other test-case, it comes with lower throughput and higher latency with respect to the number of nodes ranging from 4 to 32. Caliper toolkit allows to run the node subset that endorse particular chaincodes. The investigation shows that the proposed technique without a certificate can respond within 1 to 16 milliseconds. However, it delays 40 to 242 ms with the default CA of the Hyperledger CBC deployment. The response latency varies with the increase of workloads.

% the latest industry infrastructures largely depend on AI-driven maintenance, the prediction based on corrupted data undoubtedly results in loss of life and capital. Depending on a single trusted party, i.e., Certificate Authority (CA) indulges the trust challenges of the industry.%

% The shortcomings of the conventional cloud or trusted certificate-driven techniques have motivated us to exhibit a novel Blockchain-based AI-enabled APT detection and protection technique for Industry 4.0 Cyber-physical system. 

\section{Conclusion}

Security of the critical Industry 4.0 Cyber-physical system deserves immense concern as any leakage should outcome devastating financial damages and loss of lives. On top of malware, ransomware Advanced Persistent Threat (APT) has been responsible for such loss. Industry 4.0 edge ecosystem wonders for a cooperative trust-building rather than trusting a single entity that the proposed security technique purposely promises to offer through a certificateless mechanism.  Admittedly, an inadequate data-protection mechanism can readily challenge the security and reliability of the network. Considering the detection accuracy, the proposed approach has utilized the salient features of the deep transfer learning (DTL) algorithm upon the residual neural network (ResNet) model. After successfully filtering the APT from the edge end, that data is transferred to the associated distributed hash table storage (DHT). Consortium Blockchain (CBC) network ensures the IoT sensor registration, authentication, and validation. The immutable ledger records the data and APT detection transaction. The proposed detection model has an overall accuracy score of about 90\% where CBC increases the data transaction rate. 

%In the future, we will also concentrate on improving the system's performance and accuracy depending on the more realistic industry requirements besides investigating on the data privacy needs.

\ifCLASSOPTIONcaptionsoff
  \newpage
\fi

\bibliography{References}
\bibliographystyle{ieeetr}

\begin{IEEEbiography}[{\includegraphics[width=1in,height=1.25in,clip,keepaspectratio]{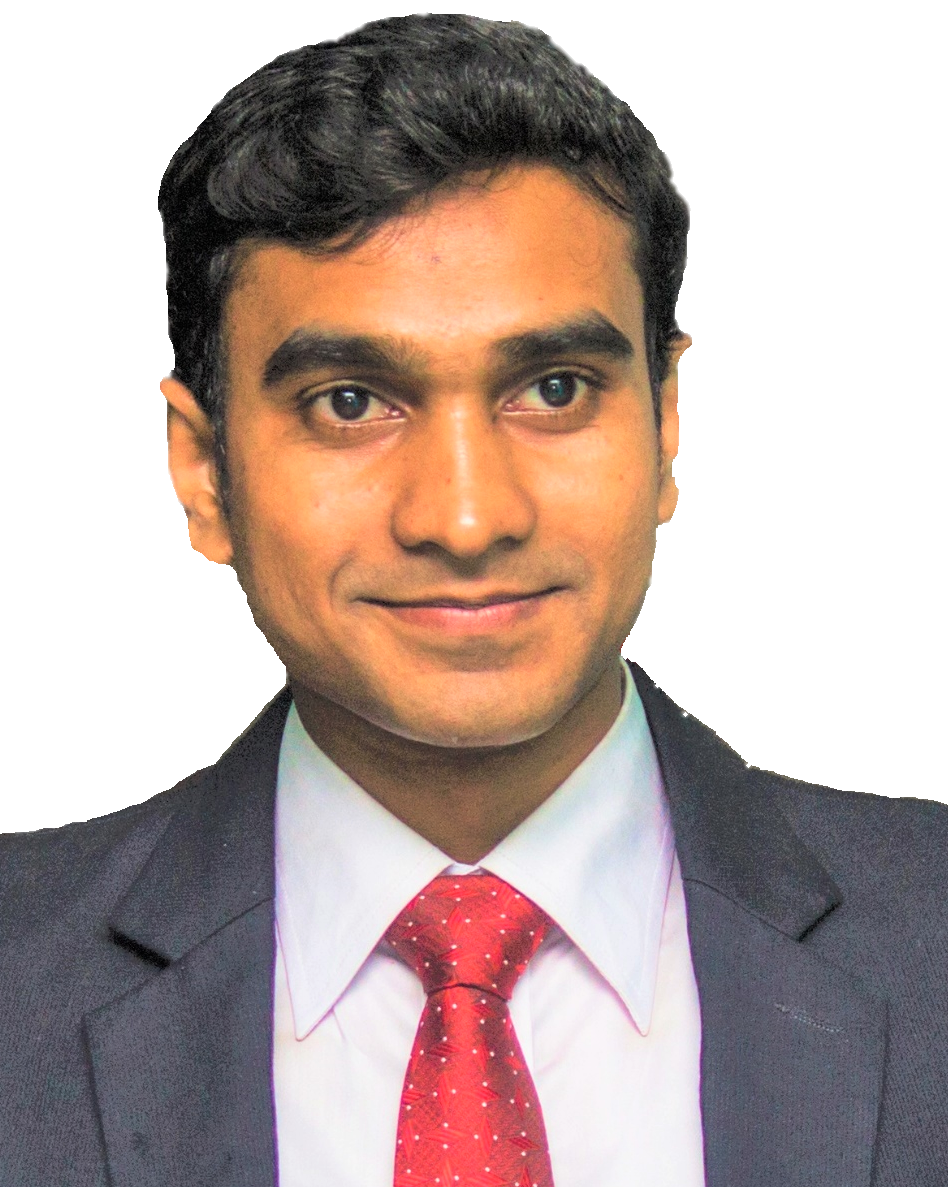}}]{Ziaur Rahman} is a doctoral research scholar in Cybersecurity of RMIT University, Melbourne. He served Mawlana Bhashani Science \& Technology University, Bangladesh as an Associate Professor in ICT. He casually served RMIT, Monash, Deakin and Charles Sturt University, Australia. Three (03) articles he coauthored were nominated and received the best paper awards. He is affiliated with the IEEE, ACM, Australian Computer Society. His research interests include blockchain technology, security of the internet of things (IoT), machine learning.
\end{IEEEbiography}

\begin{IEEEbiography}[{\includegraphics[width=1in,height=1.25in,clip,keepaspectratio]{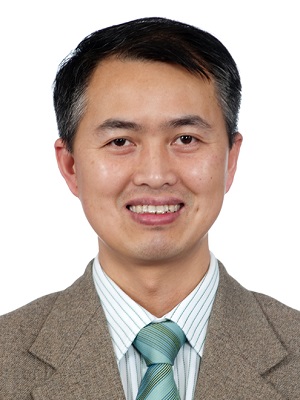}}]{Xun   Yi} is currently a full Professor of Cybersecurity with the School of Computing Technologies  with  the  School of   Science, RMIT University,   Melbourne, VIC, Australia. He has published more than 200 research papers in international journals and conference pro-ceedings. His research interests include applied cryptography, computer and network security, mobile and wireless  communication  security,  and  data  privacy protection. Prof. Yi has ever undertaken program committee members for more than 30 international conferences. Recently, he has led some Australia Research Council Discovery Projects in Data Privacy Protection. From 2014 to 2018, he was an Associate Editor for IEEE TRANSACTIONS ON DEPENDABLE AND SECURE COMPUTING.
\end{IEEEbiography}

\begin{IEEEbiography}[{\includegraphics[width=1in,height=1.25in,clip,keepaspectratio]{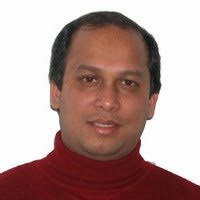}}]{Ibrahim Khalil} is currently a full  Professor  with the School of Computing Technologies,   RMIT   University,   Melbourne,  VIC, Australia. He received the Ph.D. degree from the University of Berne, Berne, Switzerland, in 2003. Before   joining   RMIT   University,   he also   worked   with   EPFL,   Lausanne,   Switzerland, University  of  Berne,  and  Osaka  University, Osaka, Japan. He has several years of experience in Silicon Valley-based  companies  working  on  large network provisioning and management software. His research interests  are  in  scalable  efficient  computing  in  dis-tributed systems, network and data security, secure data analysis, including big data security, steganography of wireless body sensor networks, and high speed sensor streams and smart grids.

\end{IEEEbiography}

\end{document}